\documentclass{caosp305}
\usepackage{graphicx}
\usepackage{natbib}
\articleNo{IBWS01}
\pubyear{2017}
\volume{47}
\volnumber{2}
\firstpage{192}
\received{May 12, 2017}
\accepted{June 5, 2017}   

\def\BibTeX{{\rm B\kern-.05em{\sc i\kern-.025em b}\kern-.08em
             T\kern-.1667em\lower.7ex\hbox{E}\kern-.125emX}}

\begin{document}

%

\hauthor{J.\,Merc, R.\,G\'alis and L.\,Leedj{\"a}rv}

\title{Recent outburst activity of the super-soft X-ray binary AG~Draconis}



\author{
        J.\,Merc \inst{1}
      \and 
        R.\,G{\'a}lis \inst{1}   
      \and 
        L.\,Leedj{\"a}rv \inst{2}
       }


\institute{
         Faculty of Science, P. J. \v{S}af{\'a}rik University,\\
         Park Angelinum 9, Ko\v{s}ice 040 01, Slovak Republic\\ \email{jaroslav.merc@student.upjs.sk}
         \and
         Tartu Observatory,\\
         Observatooriumi 1, T\~{o}ravere, 61602 Tartumaa, Estonia
          }


\date{May 12, 2017}


\maketitle


\begin{abstract}
AG~Draconis is a bright symbiotic binary consisting of a white dwarf and a pulsating cool giant. Moreover, it is the most intense X-ray source among symbiotic stars, and one of the best representatives of the super-soft X-ray objects. The system undergoes characteristic symbiotic activity with alternating quiescent and active stages. The active ones consist of several outbursts repeating at about a one-year interval. The recent activity stage of AG~Dra began with the weak pre-outburst in 2015 followed by a more prominent outburst in 2016. According to photometric and some spectroscopic observations, both brightenings belong to the minor (hot) outbursts of AG~Dra. Such behavior of the active stage is quite unusual because more often, the activity of AG~Dra starts with a major (cool) outburst. Moreover, the behavior of Raman scattered \mbox{O\,{\sc vi}} lines at $\lambda$\,6825\,\AA\, and $\lambda$\,7082\,\AA\, suggest that the minor outburst of AG~Dra in April 2016 has the characteristics of both the hot and cool outbursts. Based on the above, an open question is the next evolution of activity of the symbiotic binary AG~Dra in 2017 and beyond.

\end{abstract}

\section{AG~Draconis}
\label{sec:AGDraconis}

AG~Draconis is a classical symbiotic variable star of type S \citep{1998A&A...335..545F}. Thanks to its convenient position in the sky, its relatively high brightness (8 to 11 magnitudes in the $ V $ filter) and the low extinction, $ E (B - V) = 0.0356 \pm 0.0021 $ (based on the estimates from infrared dust maps by \citealp{2011ApJ...737..103S}), it belongs to the best-studied symbiotic systems.

The cool component of this binary is a red giant of the spectral type K3\,III \citep{2011ARep...55...31S} with an effective temperature of $T_{\rm eff} = 4\,300\,\rm K$ \citep{1998A&A...335..545F}, and mass 1.5\,M$_{\odot}$ \citep{1987AJ.....93..938K}. According to the spectral type of the giant, AG~Dra is assigned to the subclass of the so-called yellow symbiotic stars. The radius of the giant was estimated to be 33 $ \pm $ 11\,R$_{\odot}$ \citep{2005A&A...440..995S}. If we assume the radius of its Roche lobe of 170\,R$_{\odot}$ \citep{2002MNRAS.330..772O}, AG~Dra can be classified as a detached binary, and the accretion most likely takes place from the stellar wind of the cool giant.

The second component of AG~Dra is the hot white dwarf with a high temperature 1\,-\,1.5\,$\times\,10^{5}\,$K \citep{1995AJ....109.1289M,2010A&A...510A..70S} and luminosity of 1\,-\,5\,$\times\,10 ^{3} $\,L$_{\odot}$. The mass of the white dwarf was approximatively estimated as 0.4\,-\,0.6\,M$_{\odot}$. Due to the giant's wind, the binary is surrounded by an extensive circumbinary nebula, partially ionized by the white dwarf.

\begin{figure}[t]
	\includegraphics[width=\linewidth]{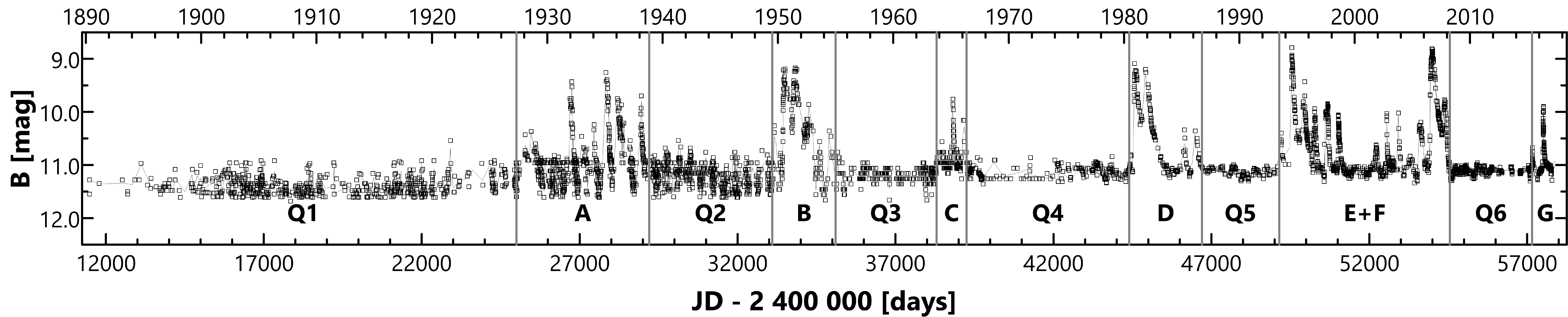}
	\caption{The historical light curve (LC) of AG~Dra over the period 1889\,-\,2017, constructed on the basis of the photographic and the $B$ band observations. The LC is divided into active (A\,-\,G) and quiescence (Q1\,-\,Q6) stages by vertical lines. The thin curve shows spline fit to the data points.
	}\label{fig:AGDrainB}
\end{figure}

There are two real periods presented in the symbiotic system AG~Dra, a~longer period of about 550 days is related to orbital motion \citep{1979IBVS.1611....1M,1999A&A...348..533G,2000AJ....120.3255F,2014MNRAS.443.1103H} and a shorter one of about 355 days is explained by the pulsation of the giant \citep{1999A&A...348..533G,2003A&A...400..595F}. 

The system manifests a characteristic symbiotic activity with the alternation of active and quiescent stages (figure \ref{fig:AGDrainB}). The individual outbursts in active stages are repeated at approximately one-year intervals. The amplitude of the outbursts is from 1\,-\,1.4\,mag in the $V$ filter to 3.6\,mag in the $U$ filter \citep{2016MNRAS.456.2558L}. The active stages occur at intervals of 9\,-\,15 years (in 1936, 1951, 1966, 1980, 1994, 2006 and 2015).

Since AG~Dra is one of the brightest symbiotic systems, many of observation programs have focused on various parts of its electromagnetic spectrum, from radio to X-ray wavelengths. AG~Dra is a source of radio emission, apparently associated with mass ejections \citep{2002MNRAS.335L..33M}. In infrared, mainly the cool component can be observed, which makes it possible to determine the physical parameters of the giant as well as to deconvolve the composite spectrum consisting of the spectra of the individual components and the envelope \citep{2005A&A...440..995S}.

In the ultraviolet region, the hot component of AG~Dra dominates and therefore it is important for determining the parameters of the compact object in this symbiotic system. For systematic research in UV, observations obtained by the IUE and FUSE have been used. \citet{1984ApJ...283..226V} and \citet{1997A&A...322..576G} found that the UV flux increased during the outbursts. \citet{1999A&A...347..478G} showed that two types of outburst are presented in AG~Dra which differ in the temperature of the hot component. Major outbursts at the beginning of active stages (e.g. 1981\,-\,1983, 1994\,-\,1996 and 2006\,-\,2008) are usually cool, during which the expanding pseudo-atmosphere of the WD cools down and the \mbox{He\,{\sc ii}} Zanstra temperature drops. In smaller scale hot outbursts, the \mbox{He\,{\sc ii}} Zanstra temperature increases or it remains unchanged. \citet{2016MNRAS.456.2558L} showed that the cool and hot outbursts of AG~Dra could be clearly distinguished by the behavior of emission lines in the optical spectrum of this symbiotic system.

AG~Dra is also a source of super-soft X-rays. Observations in this spectral region were acquired by the HEAO-2, EXOSAT, ROSAT, Chandra and XMM-Newton satellites. The observations have shown that the X-ray flux does not change during the quiescence, but it decreases significantly during the outbursts \citep{1997A&A...322..576G,2008A&A...481..725G,2009A&A...507.1531S}. It follows that the X-ray variability is in anti-correlation with the brightness changes in UV and optical.

\section{Observations}
\label{sec:Observations}

We use all photometric observations of AG~Dra that had been already analyzed and discussed in our previous study \citep{2014MNRAS.443.1103H}. New photometric data were obtained from \textit{AAVSO International Database} \citep{aavso} and \citet{vrastak}.

Intermediate-dispersion spectroscopy of AG~Dra was carried out at the Tartu Observatory in Estonia. Altogether, 515 spectra obtained during almost 14 yr (from JD~2\,450\,703.3 to JD~2\,455\,651.5) on the 1.5-metre telescope (R = 6\,000, 7\,000 and 20\,000), were analyzed in our paper \citep{2016MNRAS.456.2558L}. In the recent study, we used these measurements to compare the spectroscopic behavior of AG~Dra during the outbursts in 2015\,-\,2016 with a previous activity of this interacting binary.

New spectroscopic observations were obtained from \textit{Astronomical Ring for Access to Spectroscopy database} (ARAS). We used 183 spectra obtained between JD~2\,457\,102.4 (March 2015) and JD~2 457 867.4 (April 2017). Even the spectra were acquired with small telescopes (25\,-\,35\,cm, R = 1\,800\,-\,11\,000), they provided us valuable information about the recent activity of AG~Dra.

In our analysis, we focused on the strongest emission lines in the wavelength regions under study: the hydrogen Balmer lines \mbox{H$_{ \alpha }$} $\lambda$\,6563\,\AA\, and \mbox{H$_{ \beta }$} $\lambda$\,4861\,\AA, the neutral helium \mbox{He\,{\sc i}} line at $\lambda$\,6678\,\AA, the ionized helium \mbox{He\,{\sc ii}} line at $\lambda$\,4686\,\AA\, and the Raman scattered \mbox{O\,{\sc vi}} lines at $\lambda$\,6825\,\AA\, and $\lambda$\,7082\,\AA. We have normalized the spectra using the \textit{specnorm.py}\footnote{http://python4esac.github.io/plotting/specnorm.html} code markedly modified by \citet{eenmae} with minor changes conducted by us. Equivalent widths (EWs), fluxes in lines, peak intensities relative to the continuum and the positions of these lines were measured using \textit{ESO-MIDAS} package.

\section{Recent outburst activity}
\label{sec:Recent}

Active stages of AG~Dra consist of a series of individual outbursts that repeat approximately with the one-year period. After seven years of quiescence, following a pair of major outbursts in 2006\,-\,2008, in May 2015 the system entered the new active stage, and its brightness began to rise. The maximum brightness was achieved around JD~2\,457\,166 (10.7\,mag in $B$ and 9.6\,mag in the $V$ filter). It turned out that it was a less prominent, minor outburst. We labeled it G0, as it is likely to be a precursor of activity, as it has been seen several times in the photometric history of AG~Dra (figure \ref{fig:AGDrainB}). The new active stage was confirmed by a second, more prominent outburst (G1) in April 2016, around JD~2\,457\,517. The maximum brightness of 9.9 and 9.1\,mag in the $B$ and $V$ filters, respectively, ranks this outburst to the minor outbursts of AG~Dra.

\begin{figure}[t]
	\includegraphics[width=\linewidth]{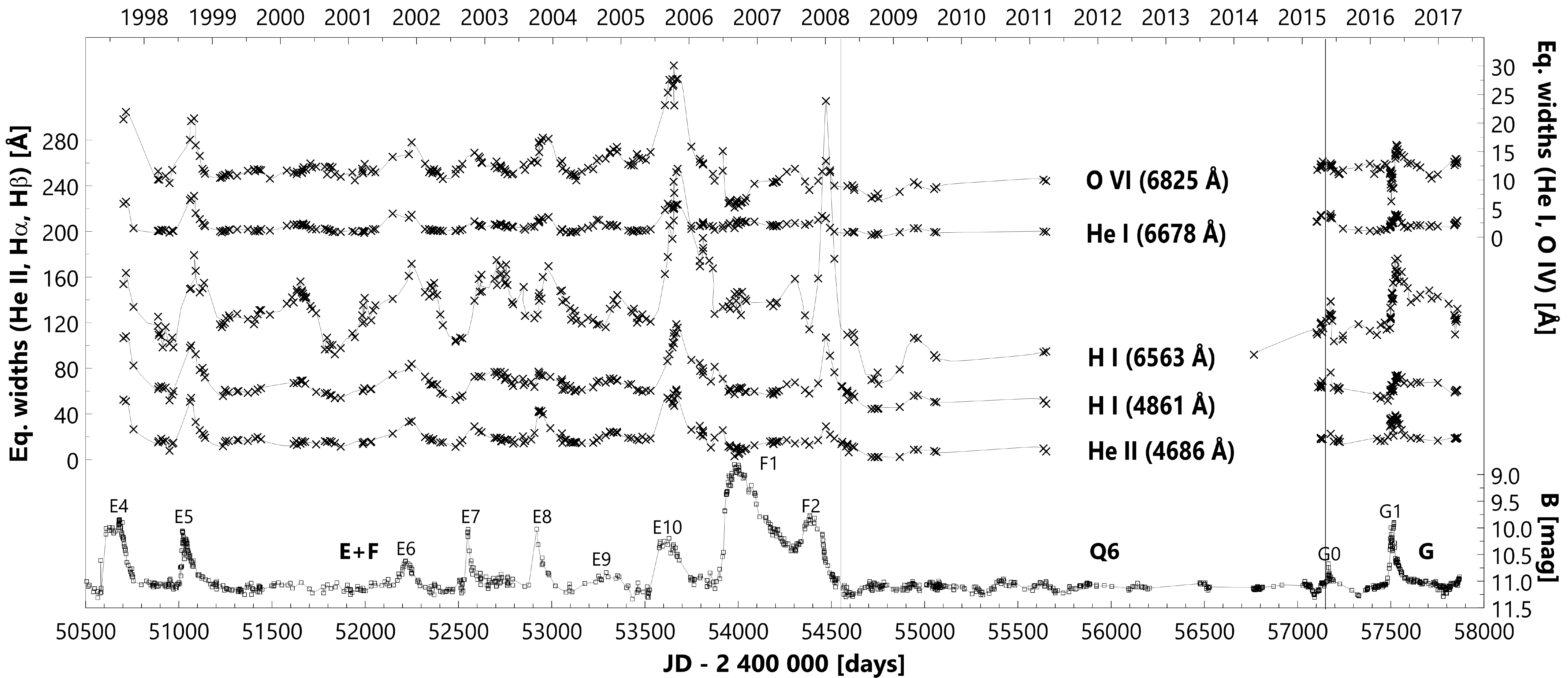}
	\caption{The curves of EWs for particular spectral lines together with the LC of AG~Dra in the $B$ filter. The scales on the left and right axes are valid for EWs of \mbox{He\,{\sc i}} $\lambda$\,6678\,\AA\, and \mbox{He\,{\sc ii}} $\lambda$\,4686\,\AA, respectively. Particular outbursts are assigned as E4\,-\,E10, F1, F2 and G0, G1. The active (E+F, G) and quiescent (Q6) stages are distinguished by the vertical lines. The thin curves show spline fits to the data points.}
	\label{fig:AGDraeqws}
\end{figure}

Such behavior is quite unusual. More often the active stages of AG~Dra started with major outbursts during which the brightness of the binary reached 8.8 and 8.4\,mag in the $B$ and $V$ filters, respectively. Before the major outbursts, precursors have usually appeared (at least for the active stages B, E, and probably C) with the brightness around 10.4 and 9.4\,mag in the $B$ and $V$ filters, respectively. The minor outburst at the beginning of the active stage was probably observed only during the activity stage C in 1963\,-\,1966, which was the shortest one. This stage is covered only by (relatively unreliable) photographic measurements.

During both outbursts (G0 and G1) of the recent active stage of AG~Dra, the increase of the EWs of \mbox{He\,{\sc ii}} $\lambda$\,4686\,\AA, \mbox{H$_{ \beta }$} $\lambda$\,4861\,\AA, \mbox{H$_{ \alpha }$} $\lambda$\,6563\,\AA\, and \mbox{He\,{\sc i}} $\lambda$\,6678\,\AA\, was observed (figure \ref{fig:AGDraeqws}). Such behavior suggests that these outbursts belong to the hot type \citep{2016MNRAS.456.2558L}. On the other hand, the EWs of the \mbox{O\,{\sc vi}} lines $\lambda$\,6825\,\AA\, and $\lambda$\,7082\,\AA\, dropped significantly during G1. We detected such behavior only during the main cool outburst in 2006 \citep{2016MNRAS.456.2558L}. As regards this particular brightening, is it a new type of outburst or some transition between (or a combination of) the hot and cool outbursts?

\subsection{Temperature of the hot component}

In \citet{2016MNRAS.456.2558L}, we analyzed the evolution of the temperature of the hot component of the AG~Dra system using the spectroscopic observations from Tartu Observatory. Based on the formula provided by \citet{1981psbs.conf..517I}, we showed that the ratio of the EWs of lines \mbox{He\,{\sc ii}}/\mbox{H\,$_{\beta}$} could be a good proxy for the temperature of the central source of ionizing photons: 

\begin{equation}
T_{hot}\,(in\,10^4\,K) \approx 14.16 \sqrt{EW_{4686} \over EW_{H_{\beta}} } + 5.13
\end{equation}

\begin{figure}[t]
	\includegraphics[width=\linewidth]{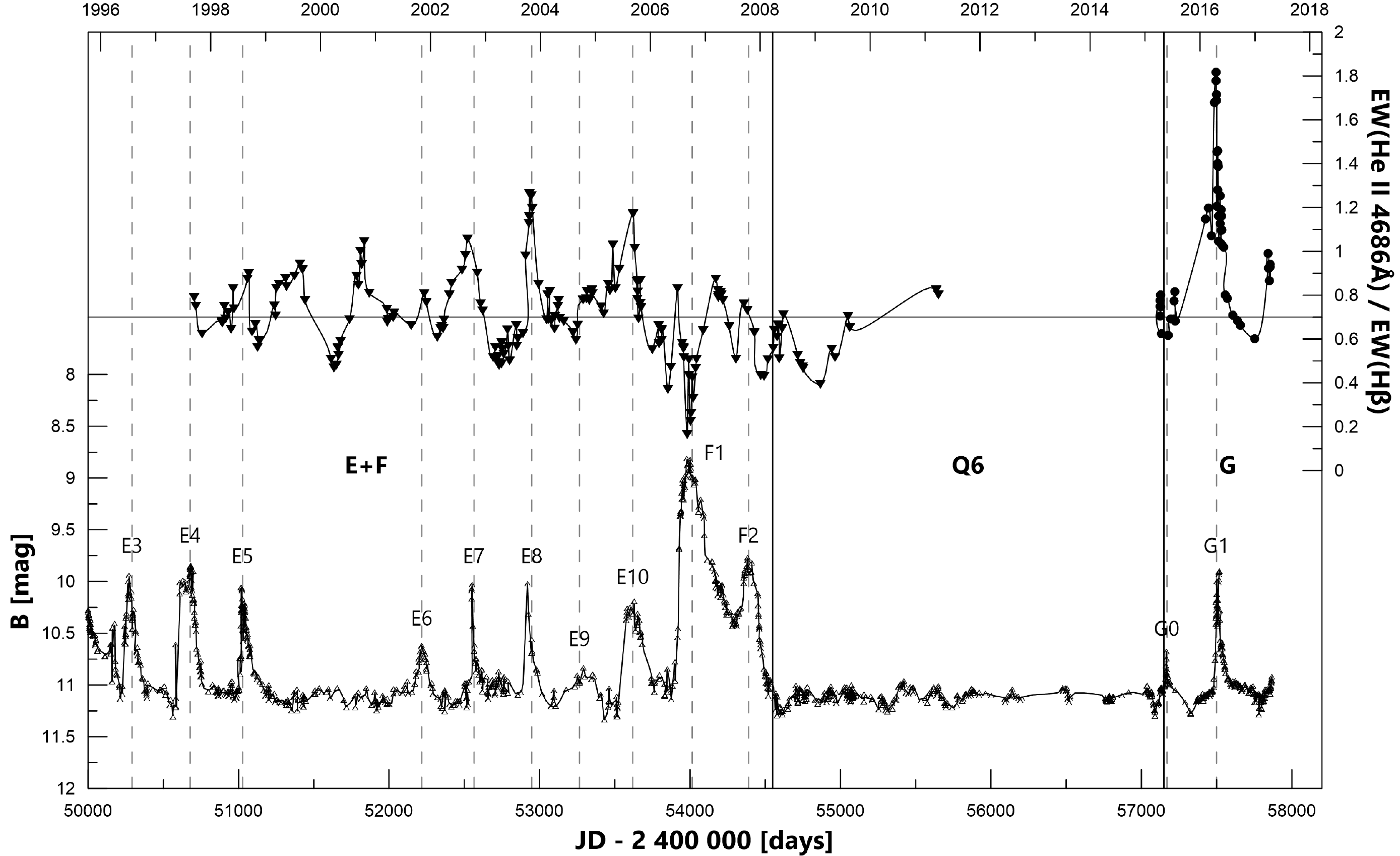}
	\caption{The ratio of the EWs of two strong emission lines H$ _{\beta} $ and \mbox{He\,{\sc ii}} $ \lambda $\,4686\,\AA, in time. The data from \citet{2016MNRAS.456.2558L} and ARAS database are depicted by triangles and circles, respectively. The stages of activity and quiescence are distinguished by the solid vertical lines. The individual outbursts are marked by the dashed vertical lines. The horizontal line corresponds to the value of 0.70 which is the average ratio of the EWs of these emission lines in the period of 1997\,-\,2011.}
	\label{fig:temperature}
\end{figure}

During the period of 1997\,-\,2011 the value of this ratio exceeded 1.0 only during the hot outbursts, and it decreased markedly during the major (cool) outburst F1 around JD~2\,454\,000. We derived that the long-term average of this ratio was 0.70 $ \pm $ 0.18, corresponding to the hot component temperature of about 170\,000\,K. The maximum value of 1.26 (210\,000\,K) and the minimum value of 0.16 (108\,000\,K) were achieved during the hot outburst E8 and the main cool outburst F1 (figure \ref{fig:temperature}), respectively. Other episodes of the low \mbox{He\,{\sc ii}}/H$ _{\beta} $ values occurred around JD~2\,454\,480 (F2) and JD~2\,454\,840 (quiescence stage Q6). The latter period corresponds to the time when the minimal fluxes of all monitored emission lines were observed. Similar drops to 0.5 were also observed around JD~2\,451\,640 (probably a short quiescence between E5 and E6) and JD~2\,452\,750 (between E7 and E8).

Similarly, we detected several periods when the ratio He\,{\sc ii}/$H_{\beta}$ reached the values $ > $1.0 (around JD~2\,451\,810 and JD~2\,452\,520) and about 1.2 around JD~2\,452\,940 and JD~2\,453\,620. The first of these events occurred during a short quiescence between E5 and E6, and the remaining three episodes corresponded to the hot outbursts (E7, E8, and E9).

During the recent active stage G, we detected noticeable changes of the He\,{\sc ii}/$H_{\beta}$ ratio. Within the minor outburst G0 (JD~2\,457\,166), the ratio corresponded to the average value of the previous period. Unfortunately, this outburst was not sufficiently covered by the spectroscopic observations to allow for a more detailed analysis. On the other hand, the G1 outburst (JD~2\,457\,517) is accompanied by an extreme temperature increase that was not recorded before. The He\,{\sc ii} and $H_{\beta}$ ratio reached the maximum of 1.82, corresponding to the temperature of 242\,000\,K. After the G1 outburst, the temperature dropped to its average level and begun to rise again at the beginning of 2017.

According to the temperature of the hot component, we can conclude that at least the G1 outburst belongs to the hot type of outbursts of AG~Dra. However, the inherent structure of the outburst (anti-correlation of the evolution of the hot component temperature and the brightness changes), as well as behavior of some emission spectral lines (disappearance of the Raman scattered lines of \mbox{O \sc vi}, see below), also demonstrate a typical evolution observed during the cool outbursts of AG~Dra. The question is whether this is a new type of activity stage or we have already observed such behavior during the weak active stage C. Unfortunately, there are no spectroscopic observations of AG~Dra in this period, so a direct comparison is not possible.
 
\subsection{Raman scattered \mbox{O\,{\sc vi}} lines}
The nature of two broad emission spectral lines in the spectrum of symbiotic stars (at 6825 and 7082 \AA\,wavelengths) has long been the subject of discussion. \citet{1989A&A...211L..31S} identified these lines as a product of Raman scattering of photons from the resonance lines of oxygen at 1032 and 1038 \AA\, on neutral hydrogen atoms. The formation of these lines requires specific physical conditions that exist almost exclusively in symbiotic stars - the presence of a hot radiation source capable of ionizing the oxygen atom five times, and enough neutral hydrogen.

Raman scattered lines of \mbox{O\,{\sc vi}} are present in the AG~Dra spectra permanently, they almost disappeared only during the cool outburst F1 in 2006 \citep{2009PASP..121.1070M,2010A&A...510A..70S,2016MNRAS.456.2558L}. The same effect was observed during the G1 outburst (figure \ref{fig:AGDraeqws}). However, the substance is likely to be different. During the outburst F1, the temperature of the hot component was probably not high enough for the formation of these emission lines. In the case of the outburst G1, a significant decrease of neutral hydrogen and/or the \mbox{O\,{\sc vi}} ions was probably due to their ionization as the hot component of AG~Dra reached the extreme value of the temperature during this period.

\section{Discussion and conclusion}
\label{sec:Discussion} 

After seven years of flat quiescence, the brightness of AG~Dra started to rise again in the first half of 2015, which turned out to be the weak pre-outburst. The beginning of the new active stage was definitively confirmed by the second, more prominent outburst in April 2016.

According to the photometric behavior, both brightenings belong to the minor outbursts of AG~Dra. Such photometric behavior of the active stage is very untypical because more often, the activity of AG~Dra starts with the major outburst. The spectroscopic observations demonstrated the increase of EWs of the studied emission lines which is typical behavior during hot outbursts of AG~Dra. We analyzed the evolution of the temperature of the hot component of AG~Dra and we detected that the outburst in 2016 was accompanied by its extreme value around 242\,000\,K. On the other hand, the almost disappearance of the Raman scattered lines of \mbox{O \sc vi} demonstrate typical evolution observed during the cool outbursts of AG~Dra. An open question remains if this was a new type of outburst or some transition between (or the combination of) the hot and cool outbursts. Another interesting challenge will be the next evolution of activity of AG~Dra in 2017 and beyond. 

\acknowledgements
We are grateful to all of the amateur astronomers, members of the ARAS group that contributed their observations to this paper. We acknowledge with thanks the variable star observations from the AAVSO International Database contributed by observers worldwide and used in this research. This study was supported by the Slovak Research and Development Agency Grant APVV-15-0458.

\bibliography{mercAGDra}

\end{document}